\documentclass[%
twocolumn,
superscriptaddress,
preprintnumbers,
nofootinbib,
 amsmath,amssymb,
 aps,
longbibliography,
notitlepage
]{revtex4-2}
\usepackage{amsmath,amssymb,bm,epsfig,color,graphicx}
\usepackage[mathscr]{eucal}
\usepackage{slashed}
\usepackage{cancel}
\usepackage[colorlinks=true,linkcolor=blue,citecolor=blue,urlcolor=blue]{hyperref}
\usepackage[caption=false]{subfig}
\usepackage{hyperref}
\usepackage{tikz,xcolor}
\usepackage{slashed}
\usepackage{comment}
\usepackage{cancel}
\usepackage{multirow}

\begin{document}

\preprint{UT-Komaba/24-10}

\author{Hirotsugu Fujii}
\email{hr-fujii@nishogakusha-u.ac.jp}
\affiliation{International Economics and Politics, Nishogakusha University,
6-16 Sanbancho, Chiyoda-ku, Tokyo 102-8336, Japan} 

\author{Kohei Fujikura}
\email{kfujikura@g.ecc.u-tokyo.ac.jp}
\affiliation{Graduate School of Arts and Sciences, University of Tokyo\\
Komaba, Meguro-ku, Tokyo 153-8902, Japan} 

\author{Yoshio Kikukawa}
\email{kikukawa@hep1.c.u-tokyo.ac.jp}
\affiliation{Graduate School of Arts and Sciences, University of Tokyo\\
Komaba, Meguro-ku, Tokyo 153-8902, Japan} 

\author{Takuya Okuda}
\email{takuya@hep1.c.u-tokyo.ac.jp}
\affiliation{Graduate School of Arts and Sciences, University of Tokyo\\
Komaba, Meguro-ku, Tokyo 153-8902, Japan}

\author{Juan W. Pedersen}
\email{juan.pedersen@riken.jp}
\affiliation{{\it RIKEN Center for Quantum Computing, Wako, Saitama, 351-0198, Japan}}

\title{Critical behavior of the Schwinger model
\\ via 
gauge-invariant variational uniform matrix product states
}
\begin{abstract}
We study the lattice Schwinger model
by combining the variational uniform matrix product state (VUMPS) algorithm with a gauge-invariant matrix product ansatz that locally enforces the Gauss law constraint.
Both the continuum and lattice versions of the Schwinger model with $\theta=\pi$ are known to exhibit first-order phase transitions for the values of the fermion mass above a critical value, where a second-order phase transition occurs.
Our algorithm enables a precise determination of the critical point in the continuum theory.
We further analyze the scaling in the simultaneous critical and continuum limits and confirm that the data collapse aligns with the Ising universality class to remarkable precision.
\end{abstract}

\maketitle

\def\thefootnote{\arabic{footnote}}
\setcounter{footnote}{0}

\tableofcontents

\section{Introduction}

The $(1+1)$-dimensional quantum electrodynamics, commonly called the Schwinger model~\cite{Schwinger:1962tn,Schwinger:1962tp}, is a simple but non-trivial gauge theory that shares with the $(1+3)$-dimensional quantum chromodynamics (QCD) several important properties such as chiral anomaly and confinement.
For this reason, the lattice-regularized Schwinger model has been an attractive testbed for Monte Carlo~\cite{Duncan:1981hv,Ranft:1982bi,Schiller:1983sj,Bender:1984qg,Potvin:1985gw,Bardeen:1998eq,Ohata:2023sqc,Ohata:2023gru}, tensor network~\cite{Byrnes:2002gj,Byrnes:2002nv,Banuls:2013jaa,Buyens:2013yza,Shimizu:2014uva,Shimizu:2014fsa,Banuls:2015sta,Buyens:2015tea,Banuls:2016lkq,Buyens:2017crb,Ercolessi:2017jbi,Funcke:2019zna,Magnifico:2019kyj,Okuda:2022hsq,Honda:2022edn,Angelides:2023bme}, and quantum~\cite{Martinez:2016yna,Muschik:2016tws,Klco:2018kyo,Kokail:2018eiw,Surace:2019dtp,Chakraborty:2020uhf,Kharzeev:2020kgc,Rajput:2021khs,Thompson:2021eze,Yamamoto:2021vxp,Honda:2021ovk,deJong:2021wsd,Honda:2021aum,Rajput:2021trn,Nguyen:2021hyk,Cheng:2022jnw,Tomiya:2022chr,Nagano:2023uaq,Ikeda:2023zil,Sakamoto:2023cxs,Farrell:2023fgd,Farrell:2024fit,Ghim:2024pxe,Kaikov:2024acw,Guo:2024tnb,Araz:2024bgg,Li:2024jlo} simulation algorithms.
While the Monte Carlo method based on path integrals is a powerful technique for simulating lattice gauge theories including lattice QCD, it suffers from a severe sign problem when the action is complex-valued due to a $\theta$-term or a chemical potential.
Approaches based on tensor networks or quantum computing can overcome this difficulty.
However, digital quantum computing is still in its infancy and its numerical precision is currently limited.

Among tensor network techniques, 
the matrix product state (MPS) ansatz and the density matrix renormalization group (DMRG) based on it are especially efficient when simulating $(1+1)$-dimensional lattice models in the Hamiltonian formulation.
In an MPS, the physical state is approximately represented using products of finite matrices. 
While for a gapped system an approximation that truncates the matrix at a finite size is guaranteed to work well due to the area law of the entanglement entropy, the MPS ansatz can also be usefully applied to critical systems by taking the size of the matrix finite but large.
For a review of MPS, see~\cite{2011AnPhy.326...96S}.

For a translationally invariant system in infinite volume, the uniform MPS (uMPS) ansatz, where all the matrices in the MPS are assumed to be identical, is even more powerful, drastically reducing the number of parameters to be optimized.
Another tensor network technique tailored to gauge theory is the implementation of the Gauss law constraint
through the gauge-invariant MPS ansatz~\cite{Buyens:2013yza}.
Gauge invariance restricts the possible form of the matrices in the MPS, further reducing the number of variational parameters.

In this paper, we combine the variational uniform matrix product state (VUMPS) algorithm~\cite{Zauner-Stauber:2017eqw} with the gauge-invariant ansatz~\cite{Buyens:2013yza} and use it to study the critical behaviors of the lattice Schwinger model in the Kogut-Susskind formulation~\cite{Kogut:1974ag}.
Since VUMPS was shown in~\cite{Zauner-Stauber:2017eqw} to produce the ground state more efficiently than other known methods,%
\footnote{Ref.~\cite{Zauner-Stauber:2017eqw} compared VUMPS with iDMRG~\cite{White:1992zz,McCulloch:2008aun}, iTEBD~\cite{Vidal:2006ofj}, and TDVP~\cite{Haegeman:2011zz,2014arXiv1408.5056H}.
In~\cite{Buyens:2013yza}, TDVP with the gauge-invariant ansatz was applied to the lattice Schwinger model.
}
we expect that the gauge-invariant VUMPS is the most powerful method to precisely obtain the ground state of the Schwinger model.

The phase structure and the critical phenomena in the Schwinger model have been the subject of much research.
By semi-classical analysis and bosonization, Coleman~\cite{Coleman:1976uz} originally predicted phase transitions at $\theta=\pi$ for the fermion mass (divided by the dimensionful coupling $g$) above some critical value $(m/g)_c$.
The existence of the phase transition was numerically confirmed by Hamer et al.~\cite{Hamer:1982mx}, who diagonalized the lattice Hamiltonian for a small lattice and located the critical point at $(m/g)_c = 0.325\pm 0.02$.
Schiller and Ranft~\cite{Schiller:1983sj} used the local Hamiltonian Monte Carlo method and obtained a comparable estimate.%
\footnote{Concretely, \cite{Schiller:1983sj} found $(m/g)_c=0.30(1)$ for $ga=1.0$ and $(m/g)_c$ $=0.31(1)$ for $ga=0.7$.
}
Later, Byrnes et al.~\cite{Byrnes:2002nv} utilized the DMRG method and found~$(m/g)_c=0.3335(2)$.%
\footnote{This result was confirmed by other works with different methods, including~\cite{Buyens:2017crb} (uMPS), \cite{Shimizu:2014uva,Shimizu:2014fsa} (Grassmann tensor renormalization group), and \cite{Ohata:2023sqc,Ohata:2023gru} (Monte Carlo method applied to the bosonized Schwinger model on the lattice).
}
Our result obtained by gauge-invariant VUMPS is $(m/g)_c = 0.333556(5)$.

In analyzing the ground state generated by the gauge-invariant VUMPS algorithm, the MPS transfer matrix is a useful quantity.
In particular, it encodes information about the first excitation or equivalently the correlation length.
We identify the critical point of the lattice Schwinger model as the point in the parameter space where the correlation length diverges.
Based on this, we determine the critical point of the continuum theory by extrapolating to the zero lattice spacing.

Moreover, we analyze the IR and UV scaling behaviors of the lattice Schwinger model by applying the data collapse method introduced in refs.~\cite{Vanhecke:2019pez,Vanhecke:2021noi} to the correlation length, the local order parameter (dynamical electric field), and the entanglement entropy.
We find that the IR scaling aligns well with the Ising universality class, while the UV scaling behavior is that of a conformal field theory with central charge $c=1$.
The successful double data collapse in the simultaneous critical and continuum limits also yields the estimates of $(m/g)_c$ consistent with the value quoted above.

This paper is organized as follows.
In Sec.~\ref{sec:lattice-Schwinger}, we review the Kogut-Susskind formulation of the lattice Schwinger model. 
In Sec.~\ref{sec:uMPS-gauge-inv}, we review the gauge-invariant uMPS ansatz for the ground state of the model.
We review the MPS transfer matrices in Sec.~\ref{sec:transfer-matrices}.
In Sec.~\ref{sec:location}, we determine the critical point by two methods, namely, through the divergence of the correlation length in Sec.~\ref{sec:critical point correlation length} and through the scaling analysis near the critical point in Sec.~\ref{sec:collapse}.
We conclude the paper with discussion in Sec.~\ref{sec:discussion}.

Note added: When this manuscript was nearly complete, a preprint~\cite{ArguelloCruz:2024xzi} with overlapping content was posted on the arXiv, reporting a result~$(m/g)_c=0.333561(4)$.

\section{Schwinger Model and uniform matrix product states}
\label{sec:Schwinger-uMPS}

In this section, we review the Schwinger model and uniform matrix product states.
The Schwinger model is defined by the action
\begin{equation}
\begin{aligned}
    S = \int \mathrm{d}^2x \Bigg[ 
    -\frac{1}{4} & F_{\mu\nu}F^{\mu\nu} 
    - \frac{g\theta}{4\pi} \epsilon^{\mu\nu}F_{\mu\nu} \\
    &+ i\bar{\psi} (\gamma^\mu \partial_\mu + igA_\mu)\psi 
    - m\bar{\psi}\psi 
    \Bigg],
\end{aligned}
\end{equation}
where $\psi$ is the Dirac fermion, the metric is $(\eta_{\mu\nu}) = \mathrm{diag}(1,-1)$, $\theta$ is the theta angle, $m$ is the fermion mass, $F_{\mu\nu}=\partial_\mu A_\nu-\partial_\nu A_\mu$, $\epsilon^{01}=-\epsilon_{01}=1$, $(\gamma^0,\gamma^1)=(\sigma_3,i\sigma_2)$, and $\bar{\psi}=\psi^\dag \gamma^0$.

Let us work in the temporal gauge $A_0=0$, where the canonical momentum conjugate to the gauge field $A^1$ is given by $\Pi=\partial_0A^1+g\theta/(2\pi)$.
The Hamiltonian is
\begin{align}
\int dx^1\Bigg[ \frac{1}{2} 
\left(\Pi - \frac{g\theta}{2\pi}\right)^2
\hspace{-1mm}
-i\bar{\psi}\gamma^1(\partial_1+igA_1)\psi+m\bar{\psi}\psi
    \Bigg].
\end{align}
This Hamiltonian is invariant under the residual time-independent gauge transformation generated by
\begin{align}
    G\equiv \partial_1F_{01}-g\bar{\psi}\gamma^0\psi.
\end{align}
A physical state must satisfy the Gauss law constraint $G|{\rm phys}\rangle =0$.

\subsection{Lattice-Regularized Schwinger Model}
\label{sec:lattice-Schwinger}

For numerical simulation, a lattice-regularized formulation is required.
Here we review such a formulation~\cite{Kogut:1974ag} following the treatment in ref.~\cite{Banks:1975gq}.

We keep the time continuous and discretize the space as $x^1 = na$, where $a$ is the lattice spacing and integers $n$ label the lattice sites.
The Dirac fermion $\psi=(\psi_e,\psi_o)^T$ in the continuum is discretized by the staggered fermion $\phi(n)$ as $\psi_e=\phi(n)/\sqrt{2a}$ for even $n$, and $\psi_o=\phi(n)/\sqrt{2a}$ for odd $n$.
The gauge field $U(n)=e^{-iagA_1(n)}$ lives in the link which connects the $n$-th and $(n+1)$-th sites.
The canonical conjugate of $U(n)$ on the lattice is defined by $L(n)=\partial_0A^1(n)/g+\theta/2\pi$.
Then (the non-trivial part of) the discretized canonical commutation relations are given by 
\begin{align}
\{\phi(m),\phi^\dag(n)\}=\delta_{mn},\
[L(m),U(n)]=\delta_{mn}U(m).
\end{align}
The lattice-regularized Hamiltonian is expressed as 
\begin{equation}
\begin{split}
    H_\theta=& \frac{g^2a}{2}\sum_n\left(L_n+\frac{\theta}{2\pi}\right)^2+m_{\rm lat}(-1)^n\phi^\dag(n)\phi(n)\\&-\frac{i}{2a}\sum_n\left(\phi(n)^\dag U_n\phi(n+1)-{\rm h.c.}\right).
\end{split}
\end{equation}

Fermionic variables can be transformed into the spin variables by the following Jordan-Wigner transformation~\cite{Jordan:1928wi},
\begin{align}
    \phi(n) = \sigma_-(n)\prod_{l<n}(-i\sigma_z(l)).
\end{align}
where $\sigma_{\pm}\equiv (\sigma_x\pm i\sigma_y)/2$.

The Hilbert space of the lattice Schwinger model is
spanned by the orthonormal basis states 
$|{\bm \kappa}\rangle  = \bigotimes_n |s_n\rangle \otimes |p_n\rangle$, where $\bm \kappa = (s_n,p_n)_{n\in\mathbb{Z}}$, $\sigma_z(n) |s_n\rangle =s_n|s_n\rangle~(s_n=\pm 1)$ and $L(n)|p_n \rangle = p_n |p_n\rangle~(p_n\in \mathbb{Z})$.
The Gauss law constraint 
is imposed on any physical state $|\text{phys}\rangle$:
\begin{equation}
    \begin{split}
    &G(n)\equiv L(n)-L(n-1) - \dfrac{\sigma_z(n)+(-1)^n}{2},\\
    &G(n)|{\rm phys}\rangle =0.\label{eq:gauss law}
    \end{split}
\end{equation}
The Kogut-Susskind Hamiltonian of the Schwinger model is given in terms of spin operators, up to irrelevant c-number terms, as 
\begin{align}
 H_\theta &= \sum_{n\in \mathbb{Z}}
 \Bigg[
 \frac{g^2a}{2}
\left(L(n)+\frac{\theta}{2\pi}\right)^2 +\frac{m_\text{lat}}{2} (-1)^n \sigma_z(n)\nonumber\\
 &\qquad + \frac{1}{2a}(\sigma_+(n)
 U(n)
 \sigma_-(n+1)+{\rm h.c.})\Bigg],
\label{eq:lattice-Hamiltonian}
\end{align}
Since $L(n)\in \mathbb{Z}$, we can always restrict ourselves to the region $0\leq \theta \leq \pi$ by a suitable redefinition of $L(n)$.
We emphasize that, in contrast to many works involving MPS and quantum simulations where the electric field~$L(n)$ is eliminated by solving the Gauss law constraint, our approach retains it and instead incorporates the constraint through an explicit MPS ansatz.

We denoted the mass parameter appearing in~(\ref{eq:lattice-Hamiltonian}) as $m_\text{lat}$. 
It was noticed in~\cite{Dempsey:2022nys} that the Hamiltonian obeys the relation $T H_\theta T^{-1} = H_{\theta+\pi}$ for the special value $m_\text{lat}=- g^2a/8$, where
the lattice translation~$T$ is defined by
\begin{equation}
T|\ldots,s_n,p_n,\ldots\rangle = |\ldots,s_{n-1},p_{n-1},\ldots\rangle \,.
\end{equation}
This property motives the definition
\begin{equation}
m\equiv m_\text{lat} + \frac{g^2a}{8} \,.  
\end{equation}
We will use $m$ rather than $m_\text{lat}$ in our later numerical analysis.

A general physical state $|\text{phys}\rangle $ can be expressed as a linear combination of $|{\bm \kappa}\rangle$ and is subject to the Gauss law constraint~\eqref{eq:gauss law}.
Since the dimension of the local Hilbert space is infinite, a truncation of the eigenvalue of the electric flux, $p_n=0,\pm 1,\pm 2,\cdots$ with $|p|\leq p_{\rm max}$ is performed.
We will discuss the justification for this truncation later.

The lattice Schwinger model defined by the Hamiltonian~(\ref{eq:lattice-Hamiltonian}) and the Gauss law constraint~(\ref{eq:gauss law}) is invariant under translation over two sites, $T^2$.
In addition, it is also invariant under $CT$ transformation for special values of $\theta = 0$ and $\theta= \pi$, where $CT$ is the product of $T$ and the naive charge conjugation operator $C$ defined by%
\footnote{Some works such as~\cite{Berruto:1999ga,Dempsey:2022nys} refer to our~$CT$ simply as charge conjugation~$C$.
Here we follow~\cite{Buyens:2013yza} for $\theta=0$ and also incorporate the case $\theta=\pi$.
}
\begin{equation}
C|s_n,p_n\rangle=|-s_n,-p_n- \theta/\pi\rangle \,.
\end{equation}
We note the relations
\begin{equation}
\begin{aligned}
&T L(n) T^{-1}= L(n+1)\,,
\ 
T U(n) T^{-1} = U(n+1) \,,\\
&
\qquad\qquad\quad
T \sigma_i(n) T^{-1}= \sigma_i(n+1) 
\end{aligned}
\end{equation}    
and
\begin{equation}
\begin{split}
    &CL(n)C^{-1} = -L(n)-
\frac{\theta}{\pi}
    \,,\
    C 
    U(n)
    C^{-1}=
    U(n)^{-1} \,,
    \\
    &C\sigma_{z}(n)C^{-1}=-\sigma_{z}\,,
    \quad
    C\sigma_{\pm}(n)C^{-1}=\sigma_{\mp}(n).
\end{split}
\end{equation}
The mean expectation value of the electric field
$\langle\Psi|(L(n)+L(n+1)+1)|\Psi\rangle/2$, where $|\Psi\rangle$ is a $T^2$-invariant ground state, is odd under the $CT$ transformation.
Therefore, it serves as a local order parameter for $CT$-breaking.

\subsection{Uniform matrix product states and the gauge-invariant ansatz}
\label{sec:uMPS-gauge-inv}

Let us first consider a general lattice model in the infinite volume, for which the total Hilbert space is the tensor product of local Hilbert spaces of the constant dimension $d$.
The uniform matrix product state (uMPS) ansatz~\cite{Buyens:2013yza} is the representation of a translationally invariant state
\begin{align}
    |\Psi\rangle = 
    \sum_{\bm K}
(\ldots    {\bm A}^{K_{-1}}   {\bm A}^{K_0} {\bm A}^{K_{+1}}\ldots )
     |\bm{K} \rangle \,,\label{eq:uMPS ansatz}
\end{align}
where $\bm{K} = (K_n)_{n\in\mathbb{Z}}$.
For fixed $K$, ${\bm A}^K=(A^K_{\beta\gamma})$ is a $D\times D$ matrix.
In $|\bm{K}\rangle = \bigotimes_{n\in\mathbb{Z}} |K_n\rangle$,
 the index $K_n\in\{1,\ldots,d\}$ labels the basis of the local Hilbert space.
Greek symbols $\beta, \gamma \in \{ 1,2,\cdots, D\}$ are referred to as virtual indices, and $D$ is called the bond dimension.
The uniformity means that the same set of matrices ${\bm A}^K$ appear for all sites $n$, and are used as variational parameters.

We assume that the ground state in any phase is invariant under $T^2$, while the combination $CT$ may or may not be preserved.
In this case, we can postulate the uMPS ansatz \eqref{eq:uMPS ansatz} by treating two neighboring sites as a single site.
Concretely, our ansatz is
\begin{equation}
 |\Psi\rangle = \sum_{\{s_n,p_n\}}
\big(\prod_{j\in\mathbb{Z}}
 {\bm A}^{s_{2j-1},p_{2j-1}}_1{\bm A}^{s_{2j},p_{2j}}_2\big)
 |\{s_n,p_n\}\rangle \,.
\end{equation}
It was shown in~\cite{Buyens:2013yza} that the Gauss law constraint can be locally solved by endowing ${\bm A}^{s,p}_{n}$ with the virtual index structure $(A^{s,p}_n)_{(q\alpha_q;r\beta_r)}$ and making the ansatz
\begin{align}
(A^{s,p}_n)_{(q\alpha_q;r\beta_r)} = (a^{s,q}_n)_{\alpha_q,\beta_r}\delta_{p,q+(s+(-1)^n)/2}\delta_{p,r}  \label{eq:gauge invariant ansatz}
\end{align}
for $n=1,2$.%
\footnote{The gauge-invariant uMPS has proven highly effective in investigating various properties of the Schwinger model, including the ground state and low-lying excitations~\cite{Buyens:2013yza,VanAcoleyen:2014suo}, thermal equilibrium properties~\cite{Buyens:2016ecr}, and real-time dynamics~\cite{Buyens:2016hhu}.}
Here, the indices $q$ and $r$ are identified with the eigenvalues of the electric field $L_j$.
Consequently, a gauge-invariant uMPS is parameterized by $D_q\times D_r$ matrices 
$(a^{s,q}_n)_{\alpha_q,\beta_r}$ ($n=1,2$) with $q$- or $r$-dependent bond dimensions so that $\alpha_q=1,\cdots,D_q$ and $\beta_r=1,\cdots,D_r$.
The total bond dimension of the gauge-invariant MPS is given by $D_{\rm tot}\equiv \sum_q D_q$, where $D_q$ vanishes for large enough $|q|$.

In the usual VUMPS algorithm~\cite{Zauner-Stauber:2017eqw}, the matrices $\bm{A}^K$ in~(\ref{eq:uMPS ansatz}) are optimized to produce the ground state.
In our gauge-invariant VUMPS, the matrices $(a^{s,q}_n)_{\alpha_q,\beta_r}$ are the variational parameters to be optimized.

The Schmidt decomposition in the present case takes the form~\cite{Buyens:2013yza}
\begin{align}
    |\Psi \rangle=\sum_{q\in\mathbb{Z}}\sum_{\alpha_q=1}^{D_q} \sqrt{\sigma_{q\alpha_q}}|\Psi_{q\alpha_q,L}\rangle \otimes|\Psi_{q\alpha_q, R}\rangle,
\end{align}
where the Schmidt coefficients $\sigma_{q\alpha_q}$ satisfy $\sum_{q,\alpha_q}\sigma_{q\alpha_q}=1$.
As reported in~\cite{Buyens:2017crb}, the Schmidt coefficients $\sigma_{q\alpha_q}$ for large $|q|$
are exponentially suppressed.
Indeed, a state with a larger $|q|$
leads to a larger energy cost when we restrict to the region $0\leq \theta\leq \pi$.
We also find the exponential suppression on the Schmidt coefficient of the large electric charge in our simulation.
In order to systematically control the truncation of the electric eigenvalues, we monitor the 
$q$-dependent Schmidt coefficients in each iteration within the VUMPS algorithm, and increase the value of $D_q$ 
if there is no Schmidt coefficient $\sigma_{q\alpha_q}$ that is negligible.

\section{Transfer Matrix and the correlation length}
\label{sec:transfer-matrices}

Let $\bm{A}^K$ be the matrices obtained by the VUMPS algorithm.
The corresponding transfer matrix 
is defined as
\begin{align}
(\mathcal{T}_A)_{(\beta_1\beta_2;\gamma_1\gamma_2)}\equiv \sum_K A^K_{\beta_1 \gamma_1}\bar{A}^K_{\beta_2\gamma_2}\,,
\label{def:transfer-matrix}
\end{align}
where the bar indicates complex conjugation.
In the following discussion, we regard $\mathcal{T}_A$ as a $D^2\times D^2$ matrix by identifying a pair of indices such as $\beta_1\beta_2$ as a single index.
We assume that $\mathcal{T}_A$ admits an eigendecomposition~\cite{Zauner-Stauber:2017eqw}
\begin{align}
\mathcal{T}_A=\sum_{i=0}^{D^2-1}\lambda_i|i)(i| \label{eq:eigendecomposition of transfer matrices}
\end{align}
with $1=\lambda_0>|\lambda_1|\geq\cdots\geq|\lambda_{D^2-1}|$, where $|i)$ and $(i|$ are $D^2$-component column and row vectors such that $(i|j)=\delta_{ij}$.

The two-point correlation function of local operators $O_1$ and $O_2$ is expressed as~\cite{Zauner:2014iea}
\begin{equation}
\begin{split}
    &\langle O_1(0) O_2(n+1)\rangle = (0|\mathcal{T}_{O_1}(\mathcal{T}_A)^{n}\mathcal{T}_{O_2}|0),\\
    &(\mathcal{T}_{O_i})_{(\beta_1\beta_2;\gamma_1\gamma_2)}\equiv \sum_{K_1,K_2}O^{K_1 K_2}_iA^{K_1}_{\beta_1\gamma_1} \bar{A}^{K_2}_{\beta_2\gamma_2},\\
    &O^{K_1 K_2}_i\equiv \langle K_2|O_i|K_1\rangle.
\end{split}
\end{equation}
Using~\eqref{eq:eigendecomposition of transfer matrices}, it can be rewritten as
\begin{align}
    \langle O_1(0)O_2(n+1)\rangle = \sum^{D^2-1}_{j=0}Z^j_{12}e^{(-\epsilon_j+i\phi_j)n},\label{eq:two-point epsilon}
\end{align}
where $\epsilon_j\equiv -\ln|\lambda_j|,~\phi_j \equiv \arg(\lambda_j)$ and $
    Z_{12}^j\equiv (0|\mathcal{T}_{O_1}|j)(j|\mathcal{T}_{O_2}|0)$.
We see that the $n$-dependence of the two-point function is expressed by a sum of exponential in the uMPS representation.
Since the $j=0$ contribution in the sum~\eqref{eq:two-point epsilon} is the product of one-point functions~$\langle O_1(0)\rangle \langle O_2(0)\rangle$, the connected contribution comes from $j\geq 1$.

A typical asymptotic behavior of the two-point function at long distances is 
\begin{align}
 |\langle O_1(0)O_2(n)\rangle|\sim n^{-\eta}e^{-n/\xi},~(n\gg 1).\label{eq:typical form of the correlation function}
\end{align}
Here, $\xi$ is the correlation length in units of the lattice spacing~$a$.  The exponent $\eta$ takes the value $1/2$ with a small correction for one spatial dimension in a deep gapped phase~\cite{ornstein1914accidental,Kennedy1991OrnsteinZernikeDI}, while it is fixed by the scaling dimension close to the critical point~\cite{Cardy:1996xt}.
Comparing the exponential dependence of $n$ in the above expression with the one in eq.~\eqref{eq:two-point epsilon}, the correlation length of the two-point function should be given by the smallest $\epsilon_j$ with $j\geq 1$ and $Z_{12}^j\neq 0$.
Therefore the longest correlation length in the lattice unit is given by $1/\epsilon_1$.

Reference~\cite{Zauner:2014iea} argued that the deviation from the exact result, arising from the finite bond dimension $D$, can be effectively parameterized by
\begin{equation}
\delta(D)\equiv \epsilon_2-\epsilon_1 \,.
\end{equation} 
We will employ $\delta(D)$ in our analysis of the critical behavior of the lattice Schwinger model.

\section{Locating the critical point}
\label{sec:location}

We now present our simulation results and use them to determine the value of the critical point $(m/g)_c$.
The theta angle is fixed at $\theta = \pi$.
With various input parameters 
$D$, $m/g$, and $ga$, we ran the gauge-invariant VUMPS simulations.
Each run produced the matrices $(a_n^{s,q})_{\alpha_q,\beta_r}$, from which we constructed the transfer matrix through~(\ref{eq:gauge invariant ansatz}) and~(\ref{def:transfer-matrix}).
We then obtained the output parameters
\begin{align}
\{\epsilon_1,\delta\} \,.
\end{align}
The continuum limit corresponds to $ga\to 0$, while
the limit $\delta\to 0$ ($D\to\infty$) can be interpreted as removing the infrared cut-off~\cite{2018PhRvX...8d1033R,Vanhecke:2019pez}.

Our simulations were performed on standard laptops and comparable platforms.

\subsection{Double extrapolations}\label{sec:critical point correlation length}

In this subsection, we determine the critical mass $m_c$ in the continuum limit via extrapolations, first in $\delta(D)$ and then in $ga$.

\begin{figure}
    \centering
    \includegraphics[width=0.7\linewidth]{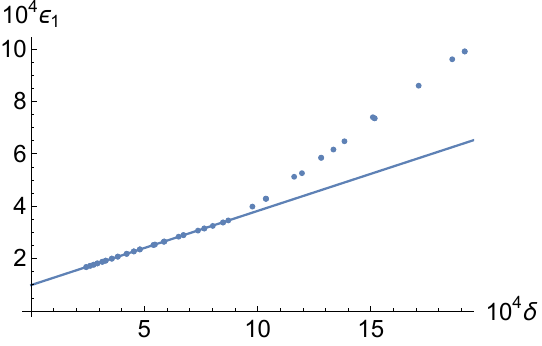}
    \caption{Plot of the simulation data for $(\delta,\epsilon_1)$ with $ga=0.1$ and $m/g=0.333$.  For $\delta<9\times 10^{-4}$ ($D>220$), $\delta$ and $\epsilon_1$ show a linear relationship as described in~(\ref{eq:e1-delta}).  The corresponding fitting line is displayed in the figure.
    For fixed $ga$ and $m/g<(m/g)_*$, the simulations with large $\delta$ (small~$D$) prepare $CT$-breaking states rather than approximating the true $CT$-preserving ground state.  See also footnote~\ref{footnote:fig2}.
    }
    \label{fig:delta-e1}
\end{figure}

\begin{figure}
    \centering
    \includegraphics[width=0.9\linewidth]{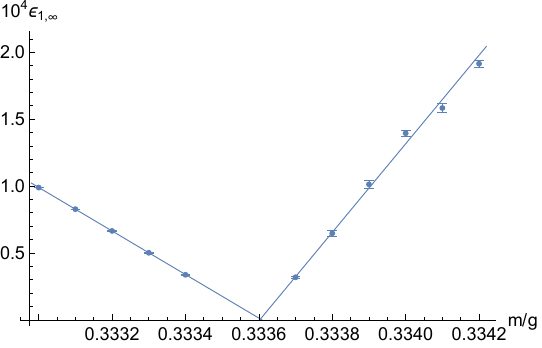}
    \caption{Plot of the data for $(m/g,\epsilon_{1,\infty})$ with $ga=0.1$.
    The data are linearly fitted separately in the two regions $m/g<0.3336$ and $m/g>0.3336$ as in~(\ref{eq:e1inf-m-two-regions}).
    Fitting in the former region gives higher precision because the data points there align more closely with a straight line and have smaller error bars.
    }
    \label{fig:mass-e1}
\end{figure}

Let us schematically describe the relations between the relevant quantities.
For fixed values of $(ga \neq 0 ,m/g)$ and for large bond dimensions $D$, the two quantities $\epsilon_1$~and $\delta$ (both dependent on $D$) are approximately related as%
\footnote{%
The assumption behind the ansatz~(\ref{eq:e1-delta}) is that the finite bond dimension effect 
is similar to the finite size effect and discretizes the continuous spectrum. 
The quantity~$\delta$ is a measure for how well the discrete spectrum approximates the exact continuous one~\cite{2018PhRvX...8d1033R,Vanhecke:2019pez}.
}
\begin{equation}\label{eq:e1-delta}
\epsilon_1(D) = \epsilon_{1,\infty}+ c_1 \delta(D)
\end{equation}
with $D$-independent parameters $ \epsilon_{1,\infty}$ and $c_1$ (FIG.~\ref{fig:delta-e1}).
The parameter $\epsilon_{1,\infty}=\epsilon_{1,\infty}(m/g)$ depends on $m/g$ as shown in FIG.~\ref{fig:mass-e1}.%
\footnote{\label{footnote:fig2}In FIG.~\ref{fig:mass-e1}, the data points at $m/g=0.3335$ and $0.3336$, which should be in the $CT$-unbroken phase according to the fitting lines for the other data points, are omitted.
This is because for these values of $m/g$ the bond dimensions $D$ up to $500$ in our simulations are too small to generate a $CT$-invariant ground state, failing to show a straight line corresponding to the left half of the plot in FIG.~\ref{fig:delta-e1}.
}
We recall from~\cite{Hamer:1982mx} that the lattice Schwinger model exhibits, even for a non-zero value of $ga$, a second-order phase transition as we vary $m/g$.  
Since $\epsilon_{1,\infty}$ is inversely proportional to the correlation length, it vanishes approximately linearly as $m/g$ approaches a ($ga$-dependent) critical value $(m/g)_*$:
\begin{equation}
\label{eq:e1inf-m-two-regions}
\epsilon_{1,\infty} = 
\left\{
\begin{array}{cc}
c_-(m_*-m)/g & \text{for } m < m_* \,,
\\
c_+(m-m_*)/g & \text{for } m > m_* \,,
\end{array}
\right.
\end{equation}
where $c_-$ and $c_+$ are $m/g$-independent positive parameters.
The combination $CT$ of charge conjugation $C$ and translation $T$ is unbroken for $m<m_*$ and broken for $m>m_*$~\cite{Hamer:1982mx}, as can be confirmed from the expectation values of the local order parameter discussed at the end of Sec.~\ref{sec:lattice-Schwinger}.
The critical mass $m_c$ in the continuum limit can be obtained as a limit
\begin{equation}
(m/g)_c = \lim_{ga\rightarrow0} (m/g)_* \,.
\end{equation}

\begin{figure}
    \centering
    \includegraphics[width=0.9\linewidth]{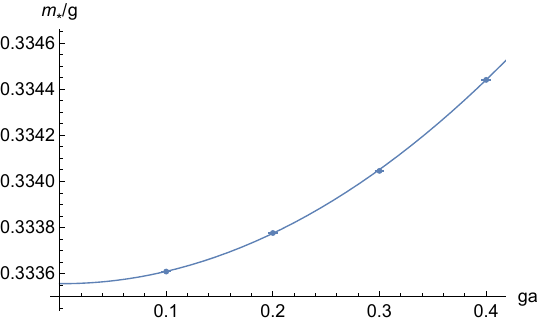}
    \caption{Plot of the data for $(ga,(m/g)_*)$, which are quadratically fitted to obtain the critical mass $(m/g)_c=\lim_{ga\rightarrow 0} (m/g)_*$ in the continuum limit.}
    \label{fig:ga-m/g}
\end{figure}

To numerically estimate $\epsilon_{1,\infty} =\epsilon_{1,\infty}(m/g,ga)$, $(m/g)_*$ $= m_*(ga)/g$, $(m/g)_c$,  and their uncertainties, we proceed as follows.
Let us take $G\equiv\{0.1,0.2,0.3,0.4\}$.
For each $ga\in G$, we choose a set $V_{ga}$ of real numbers and we run VUMPS simulations for $m/g\in V_{ga}$ for the bond dimension $D$ large enough but not exceeding 500.
For each $(ga,m/g)$, we fit the numerical data representing $(\delta,\epsilon_1)$, obtained for high enough $D$, by the linear function~(\ref{eq:e1-delta}) to obtain the value of $\epsilon_{1,\infty}$.%
\footnote{It is important to use $\delta(D)$ rather than $1/D$ for extrapolation.
For example, the linear fit with $\delta(D)$ gives a more precise estimate of $\epsilon_{1,\infty}$ than the quadratic fit with $1/D$ by a factor of 10 for the data shown in FIG.~\ref{fig:delta-e1}.
}
In the standard linear fit where the data $(x_i,y_i)$ are fitted by a linear function $y=ax+b$, the uncertainty in $b$ can be estimated from the knowledge of the uncertainty (assumed uniform, i.e., independent of $i$) in the data $y_i$, and the value of the reduced chi-squared should be close to 1 for a good fit.
In our case, we do not know the uncertainty in the value of $\epsilon_1$ for given $\delta$.
We apply the common procedure~\cite{10.5555/1403886} where we choose the uncertainty in $\epsilon_1$ such that the reduced chi-squared is 1, and use it to estimate the uncertainty in $\epsilon_{1,\infty}$.
The uncertainties are shown as error bars in FIG.~\ref{fig:mass-e1}.
We then fit the data set $\{ (m/g,\epsilon_{1,\infty}(m/g))\,|\,m/g\in V_{ga}\}$ by a linear function in $m/g$ in each of the two regions $m/g < (m/g)_*$ and $m/g>(m/g)_*$.%
\footnote{%
Since we perform VUMPS only for a finite set of the values of $m/g$, when dividing the range of $m/g$ into the two regions, it is enough to know the value of $(m/g)_*$ only roughly and its $V$-dependence can be neglected.
}
This fit gives the value of $m_{*}/g$ together with its uncertainty, for a given value of $ga$ and a choice of $V_{ga}$ (FIG.~\ref{fig:mass-e1}).
Let us denote the tuple $(V_{ga=0.1},V_{0.2},V_{0.3},V_{0.4})$ by $V$.
In the next step, we extrapolate to $ga=0$ by fitting $(m/g)_*$ with a polynomial in $ga$ and obtain the $V$-dependent estimate $m_{c,V}/g$ for the continuum critical mass and its uncertainty (FIG.~\ref{fig:ga-m/g}).

\begin{figure}
    \centering    \includegraphics[width=1\linewidth]{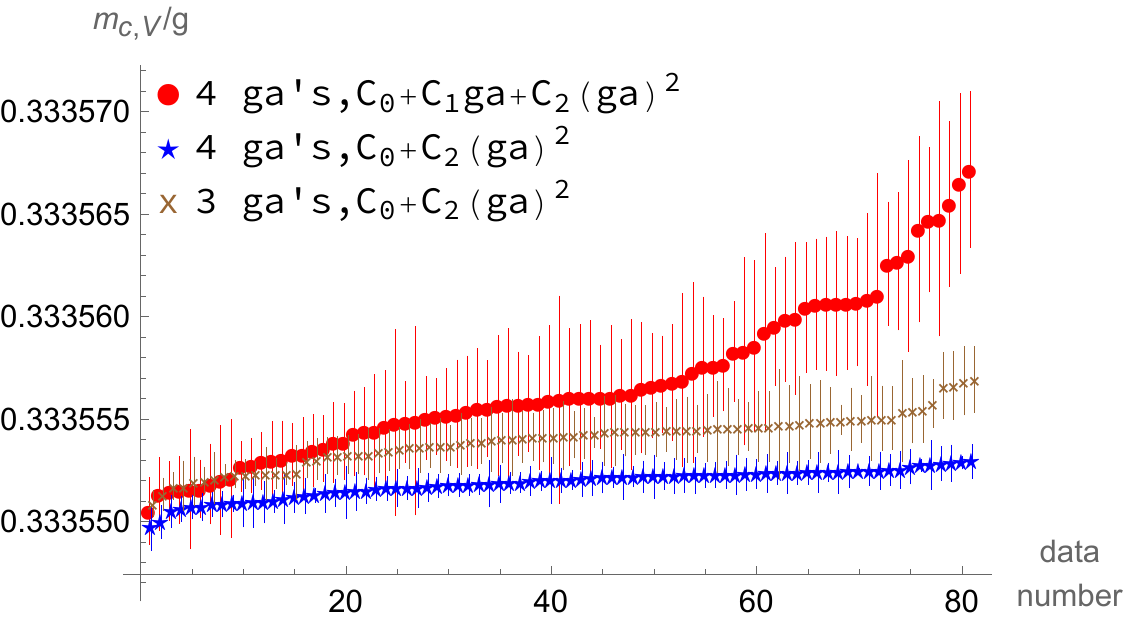}
    \caption{Sorted values of $m_{c,V}/g$ computed for random choices of $V$ and two choices of fitting polynomials.
    In each data sequence, different data numbers correspond to different choices of $V$.
    The legend for for each  sequence indicates the number of the values of $ga$ used to compute $m_{c,V}$ and the fitting polynomial.
    }
    \label{fig:dist_mcs}
\end{figure}

To quantify the sensitivity of $m_{c,V}/g$ to $V$ and the fitting method, we compute $m_{c,V}/g$ for many random choices%
\footnote{%
We fix a set $V^{(0)}_{ga}$ of real numbers, for example $V^{(0)}_{ga=0.1} = \{0.333$, $0.3331$, $0.3332$, $0.3333$, $0.3334\}$, and uniformly sample $V_{ga}$ from the set $\left\{V_{ga}\subseteq V_{ga}^{(0)} \, \middle| \, |V_{ga}|\geq 3\right\}$.}
of $V$ and two choices of the polynomial.
We then choose ``good fits" of $(ga,(m/g)_*)$, which we define to be those with the reduced chi-squared between 0.8 and 1.2.
The resulting samples of $m_{c,V}/g$ are displayed in FIG.~\ref{fig:dist_mcs}, with the error bars indicating the uncertainties obtained by fitting~\cite{10.5555/1403886}.
The red-circle data points are obtained by fitting with a polynomial $C_0+C_1 ga+C_2 (ga)^2$ the data for $(ga,(m/g)_*)$ obtained for $ga=0.1,0.2,0.3,0.4$ and various choices of $V$.
For the estimate of $(m/g)_c$, we take the median value $M$ in this sample.
As the estimate of the uncertainty in $(m/g)_c$, we take the sum of the median distance from $M$ and the median value of the uncertainty in $m_{c,V}$.
Our final estimate for the continuum critical mass with uncertainty is
\begin{equation}
(m/g)_c = 0.333556(5) \,.  \label{eq:critical point}
\end{equation}

The above fits show that the coefficient $C_1$ of $ga$ is rather small, as was originally observed in~\cite{Dempsey:2022nys} for a simulation with the mass shift (10).  This motivates us to fit the data for $(ga,m_*/g)$ with $C_0+C_2 (ga)^2$.
This gives the blue-star ($ga=0.1,0.2,0.3,0.4$) and brown-cross ($ga=0.1,0.2,0.3$) data points shown in FIG.~\ref{fig:dist_mcs}.
The procedure in the previous paragraph gives estimates for $(m/g)_c$ as $0.333552(1)$ and $0.333554(2)$ for the data sets, respectively.
These values are smaller than but consistent with~(\ref{eq:critical point}).

\subsection{Double collapse: A finite size scaling}
\label{sec:collapse}

In this subsection, we analyze the scaling behavior in the simultaneous critical and continuum limits of the lattice Schwinger model and achieve a double data collapse similar to the one observed for a different model in~\cite{Vanhecke:2019pez,Vanhecke:2021noi}.
This provides further confirmation of our estimate~(\ref{eq:critical point}).

Let us assume that the critical point of the Schwinger model is described by a scale invariant theory.
There exist the system size $L$ and the UV cutoff $\Lambda$ associated with the truncation of the bond dimension and the finite lattice spacing~\cite{2018PhRvX...8d1033R}.
Some physical observables such as the correlation length in the lattice unit $\xi$, the local order parameter (local operator that is odd under $CT$) $\phi$, and the entanglement entropy $S$ also depend on these variables, 
\begin{align}
    &\xi=\xi[t,L,\Lambda],\\
    &\phi=\phi[t,L,\Lambda],\\
    &S=S[t,L,\Lambda].
\end{align}
In these expressions, $t$ is a deviation of the relevant coupling constant from a fixed point value.
Transformation properties of these variables under scale transformations for the system size and the UV cutoff,
\begin{align}
L\to \alpha L,~\Lambda\to \Lambda/\alpha',\label{eq:scale transformation}
\end{align}
are assumed to be~\cite{Vanhecke:2021noi}
\begin{equation}
\begin{split}
    &\xi \to \alpha\xi/\alpha',~t\to \alpha^{-\Delta_t^{\rm IR}}\alpha'^{-\Delta_t^{\rm UV}}t,\\
    &\phi\to \alpha^{-\Delta^{\rm IR}_\phi}\alpha'^{-\Delta^{\rm UV}_\phi}\phi.
\end{split}
\end{equation}
Here $\Delta_t^{\rm IR(UV)}$ and $\Delta_\phi^{\rm IR(UV)}$ are the critical exponents.
For the entanglement entropy, we assume the following scaling property,
\begin{align}
    S\simeq \tilde{S}+\frac{c_{\rm UV}}{6}\log(\Lambda) + \frac{c_{\rm IR}}{6}\log(L),
\label{eq:S-St}
\end{align}
where $c_{\rm IR(UV)}$ is the central charge of conformal theories IR (UV) energy scales, and $\tilde{S}$ is
a function invariant under the scale transformations~\eqref{eq:scale transformation}.
A contribution from a non-universal part is omitted.
The scaling property with respect to $\Lambda$ is determined using the CFT formula presented in~\cite{Calabrese:2004eu}, where $\Lambda$ is identified as the inverse of the lattice spacing.

Then scale invariant functions are summarized as
\begin{align}
    &\tilde{\xi}[tL^{\Delta^{\rm IR}_t}\Lambda^{-\Delta^{\rm UV}_t}]=\xi/(\Lambda L),\nonumber \\
    &\tilde{\phi}[tL^{\Delta^{\rm IR}_t}\Lambda^{-\Delta^{\rm UV}_t}]=L^{\Delta^{\rm IR}_\phi}\Lambda^{-\Delta^{\rm UV}_\phi}\phi,\label{eq:scale invariant functions}\\
    &\tilde{S}[tL^{\Delta^{\rm IR}_t}\Lambda^{-\Delta^{\rm UV}_t}]=S-\frac{c_{\rm IR}}{6}\log(L)-\frac{c_{\rm UV}}{6}\log(\Lambda).\nonumber
\end{align}
Therefore, a randomly generated data set collapses into a single curve if all the data points are sufficiently close to the critical point.

Let us now identify 
\begin{equation}
\begin{aligned}
L&= \frac{ga}{\delta} \,,
\quad
\Lambda= \frac{1}{ga} \,,
\quad
t=\frac{m}{g}-\left(\frac{m}{g}\right)_* \,, 
\\
\phi&=\langle\Psi|(L(n)+L(n+1)+1)|\Psi\rangle/2,
\quad
\xi=\frac{1}{\epsilon_1} \,,
\end{aligned}    
\end{equation}
 where $(m/g)_*$ is the critical point of the lattice Schwinger model.
$(m/g)_*$ is not universal and hence it is non-trivial function of $\delta$ and $ga$.
We use the following ansatz of the critical point: 
\begin{align}
 (m/g)_* 
= (m/g)_c+b_1 (ga)+b_2 (ga)^2 +l_1\frac{\delta}{ga}.
\end{align}
There can be higher-order terms of $ga$ and $\delta/(ga)$, but this ansatz works well in practice, at least for small lattice spacing $0.075\lesssim ga\lesssim 0.5$ and large bond dimension $D\gtrsim 50$.

We assume the Ising universality class, $\Delta^{\rm IR}_{t}=1,~\Delta_{\phi}=1/8,~c_{\rm IR}=1/2$, for the IR scale transformation, and  $\Delta^{\rm UV}_{t}=0,~\Delta_{\phi}=0,~c_{\rm UV}=1$ for the UV scale transformation.
The remaining parameters $(m/g)_c,~b_1$ and $b_2$ should be optimized such that randomly generated data points are aligned into a single curve.
Notice that nonzero $l_1$ yields universal shifts of $\tilde{\xi},~\tilde{\phi}$ and $\tilde{S}$.

In our analysis, we take as the cost function the total length of all data points defined by
\begin{equation}
\begin{aligned}
&\quad\    F_{\tilde{f}}[(m/g)_c,~\{b_i\}]
\\
&\equiv \sum_I\sqrt{(\tilde{f}_I-\tilde{f}_{I-1})^2+(t_IL_I-t_{I-1}L_{I-1})^2},
\end{aligned}
\end{equation}
where $\tilde{f}\equiv \tilde{\xi},~\tilde{\phi},~{\rm or}~\tilde{S}$, and the index $I$ represents an ordered data point such that $\cdots<t_{I-1}L_{I-1}<t_IL_I<\cdots$. 
The minimization problem is numerically solved by the gradient descent method.
Due to the presence of many local minima near the global minimum, we solve the minimization problem multiple times with different initial guesses.

We randomly and uniformly generate 1,780 data points near the critical point within the ranges  $0.332<m/g<0.334$,  $50<D<520$, and $0.075<ga<0.4$.
The best-optimized values to achieve successful data collapse are determined for the correlation length, the order parameter, and the entanglement entropy as follows:
\begin{align}
&\tilde{\xi}:(m/g)_c=0.333560,\nonumber\\
&\tilde{\phi}:(m/g)_c=0.333560,\\
&\tilde{S}:(m/g)_c=0.333559.\nonumber
\end{align}
All the values are consistent with \eqref{eq:critical point}.
\begin{figure*}[t!]
    \centering
    
    \includegraphics[width=0.4\textwidth]{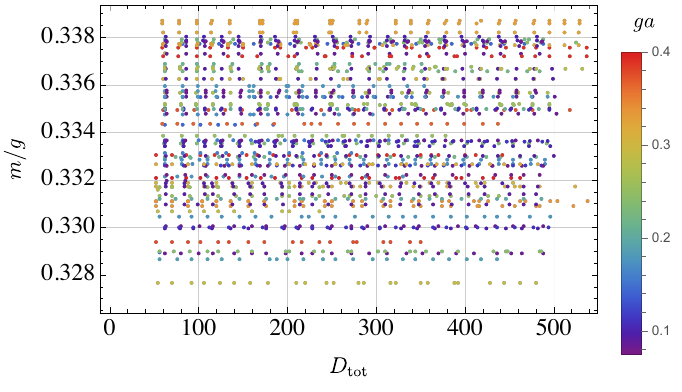} \hspace{5mm}  \includegraphics[width=0.4\textwidth]{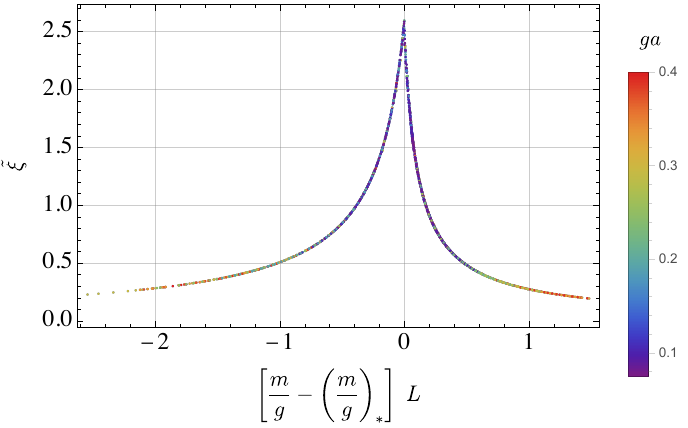}
    \includegraphics[width=0.4\textwidth]{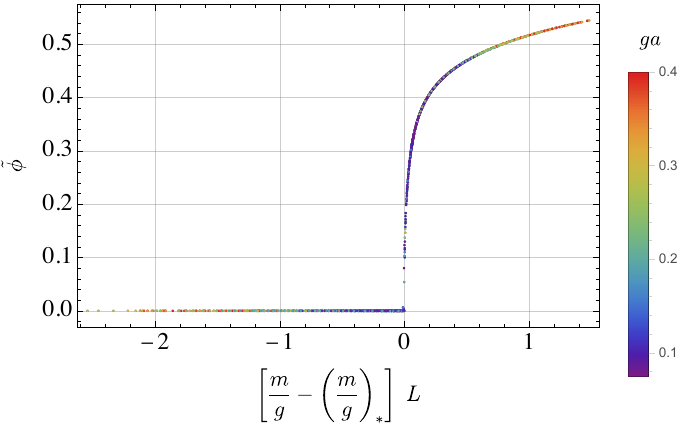} \hspace{5mm} \includegraphics[width=0.4\textwidth]{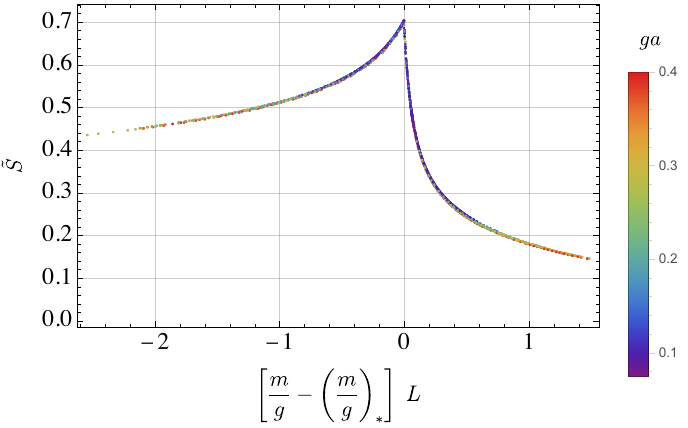}

    \caption{Color online. Generated data points used in the double collapse on $(m/g,D_{\rm tot})$-plane (top left). Double collapse of the correlation length (top right), the local order parameter (bottom left) and the entanglement entropy (bottom right). The UV and IR exponents are fixed. See the main text for details.}
\label{fig:double_collapse}
\end{figure*}
In FIG.~\ref{fig:double_collapse},
we plot the three scale invariant functions in eq.~\eqref{eq:scale invariant functions}, each with the corresponding optimized parameter set.
A distinct color is assigned to each $ga$ value, 
and the generated data points on the $(m/g, D_{\rm tot})$-plane are also shown.
One can see from each panel that all data points collapse onto a single curve.
The overall behavior of double collapse in FIG.~\ref{fig:double_collapse} is the same as that of the Euclidean $\lambda \phi^4$-theory~\cite{Vanhecke:2019pez,Vanhecke:2021noi}.

\section{Conclusion and Discussion}
\label{sec:discussion}

We studied the critical behavior of the lattice Schwinger model in the Kogut-Susskind formulation using the uMPS representation of the ground state.
The ground state was obtained by the VUMPS algorithm applied to the special uMPS ansatz where all the variational degrees of freedom are restricted to the gauge-invariant subspace.
We determined the precise value of the critical mass in the continuum Schwinger model as $(m/g)_c=0.333556(5)$. 
We also demonstrated the double collapse of the randomly generated numerical values for several physical quantities
around the critical point (FIG.~\ref{fig:double_collapse}), and confirmed that the IR critical exponents are consistent with those of the Ising universality class.

It would be interesting to apply the gauge-invariant VUMPS to a multi-flavor Schwinger model, which has a richer phase structure~\cite{Coleman:1976uz,Gepner:1984au,Affleck:1985wa,Dempsey:2023gib}, and non-abelian gauge theories such as the $(1+1)$-dimensional adjoint QCD~\cite{Dempsey:2023fvm}.

\vspace{4mm}
\acknowledgments
K.\,F. and T.\,O. thank the participants of the workshop TENSOR NETWORK 2024, where preliminary results were presented, for useful comments and discussions.
The research of T.\,O. was supported in part by Grant-in-Aid for Transformative Research Areas (A) ``Extreme Universe'' No.\,21H05190 and by JST PRESTO Grant Number JPMJPR23F3.
K.\,F. is supported
by JSPS Grant-in-Aid for Research Fellows Grant No.22J00345.
H.\,F.'s work is partially supported by Grant-in-Aid, Kakenhi 21K03568.
The work of J.\,W.\,P. is supported in part by the JSPS Grant-in-Aid, Kakenhi 24K22889.

\bibliographystyle{JHEP}
\bibliography{refs}

\providecommand{\href}[2]{#2}\begingroup\raggedright\begin{thebibliography}{10}

\bibitem{Schwinger:1962tn}
J.~S. Schwinger, \emph{{Gauge Invariance and Mass}}, \href{https://doi.org/10.1103/PhysRev.125.397}{\emph{Phys. Rev.} {\bfseries 125} (1962) 397}.

\bibitem{Schwinger:1962tp}
J.~S. Schwinger, \emph{{Gauge Invariance and Mass. 2.}}, \href{https://doi.org/10.1103/PhysRev.128.2425}{\emph{Phys. Rev.} {\bfseries 128} (1962) 2425}.

\bibitem{Duncan:1981hv}
A.~Duncan and M.~Furman, \emph{{Monte Carlo Calculations With Fermions: The Schwinger Model}}, \href{https://doi.org/10.1016/0550-3213(81)90050-X}{\emph{Nucl. Phys. B} {\bfseries 190} (1981) 767}.

\bibitem{Ranft:1982bi}
J.~Ranft and A.~Schiller, \emph{{Local Hamiltonian Monte Carlo Study of the Massive Schwinger Model in an External Background Field}}, \href{https://doi.org/10.1016/0370-2693(83)91591-5}{\emph{Phys. Lett. B} {\bfseries 122} (1983) 403}.

\bibitem{Schiller:1983sj}
A.~Schiller and J.~Ranft, \emph{{The Massive Schwinger Model on the Lattice Studied via a Local Hamiltonian Monte Carlo Method}}, \href{https://doi.org/10.1016/0550-3213(83)90049-4}{\emph{Nucl. Phys. B} {\bfseries 225} (1983) 204}.

\bibitem{Bender:1984qg}
I.~Bender, H.~J. Rothe and K.~D. Rothe, \emph{{Monte Carlo Study of Screening Versus Confinement in the Massless and Massive Schwinger Model}}, \href{https://doi.org/10.1016/S0550-3213(85)80006-7}{\emph{Nucl. Phys. B} {\bfseries 251} (1985) 745}.

\bibitem{Potvin:1985gw}
J.~Potvin, \emph{{A nonperturbative study of hadronization with heavy sources. 1. The screening length as a function of the quark mass in the Schwinger model}}, \href{https://doi.org/10.1103/PhysRevD.32.2070}{\emph{Phys. Rev. D} {\bfseries 32} (1985) 2070}.

\bibitem{Bardeen:1998eq}
W.~A. Bardeen, A.~Duncan, E.~Eichten and H.~Thacker, \emph{{Quenched approximation artifacts: A Study in two-dimensional QED}}, \href{https://doi.org/10.1103/PhysRevD.57.3890}{\emph{Phys. Rev. D} {\bfseries 57} (1998) 3890}.

\bibitem{Ohata:2023sqc}
H.~Ohata, \emph{{Monte Carlo study of Schwinger model without the sign problem}}, \href{https://doi.org/10.1007/JHEP12(2023)007}{\emph{JHEP} {\bfseries 12} (2023) 007} [\href{https://arxiv.org/abs/2303.05481}{{\ttfamily 2303.05481}}].

\bibitem{Ohata:2023gru}
H.~Ohata, \emph{{Phase diagram near the quantum critical point in Schwinger model at~$\theta = \pi$: analogy with quantum Ising chain}}, \href{https://doi.org/10.1093/ptep/ptad151}{\emph{PTEP} {\bfseries 2024} (2024) 013B02} [\href{https://arxiv.org/abs/2311.04738}{{\ttfamily 2311.04738}}].

\bibitem{Byrnes:2002gj}
T.~Byrnes, P.~Sriganesh, R.~J. Bursill and C.~J. Hamer, \emph{{Density matrix renormalization group approach to the massive Schwinger model}}, \href{https://doi.org/10.1016/S0920-5632(02)01416-0}{\emph{Nucl. Phys. B Proc. Suppl.} {\bfseries 109} (2002) 202} [\href{https://arxiv.org/abs/hep-lat/0201007}{{\ttfamily hep-lat/0201007}}].

\bibitem{Byrnes:2002nv}
T.~Byrnes, P.~Sriganesh, R.~J. Bursill and C.~J. Hamer, \emph{{Density matrix renormalization group approach to the massive Schwinger model}}, \href{https://doi.org/10.1103/PhysRevD.66.013002}{\emph{Phys. Rev. D} {\bfseries 66} (2002) 013002} [\href{https://arxiv.org/abs/hep-lat/0202014}{{\ttfamily hep-lat/0202014}}].

\bibitem{Banuls:2013jaa}
M.~C. Ba\~nuls, K.~Cichy, K.~Jansen and J.~I. Cirac, \emph{{The mass spectrum of the Schwinger model with Matrix Product States}}, \href{https://doi.org/10.1007/JHEP11(2013)158}{\emph{JHEP} {\bfseries 11} (2013) 158} [\href{https://arxiv.org/abs/1305.3765}{{\ttfamily 1305.3765}}].

\bibitem{Buyens:2013yza}
B.~Buyens, J.~Haegeman, K.~Van~Acoleyen, H.~Verschelde and F.~Verstraete, \emph{{Matrix product states for gauge field theories}}, \href{https://doi.org/10.1103/PhysRevLett.113.091601}{\emph{Phys. Rev. Lett.} {\bfseries 113} (2014) 091601} [\href{https://arxiv.org/abs/1312.6654}{{\ttfamily 1312.6654}}].

\bibitem{Shimizu:2014uva}
Y.~Shimizu and Y.~Kuramashi, \emph{{Grassmann tensor renormalization group approach to one-flavor lattice Schwinger model}}, \href{https://doi.org/10.1103/PhysRevD.90.014508}{\emph{Phys. Rev. D} {\bfseries 90} (2014) 014508} [\href{https://arxiv.org/abs/1403.0642}{{\ttfamily 1403.0642}}].

\bibitem{Shimizu:2014fsa}
Y.~Shimizu and Y.~Kuramashi, \emph{{Critical behavior of the lattice Schwinger model with a topological term at $\theta=\pi$ using the Grassmann tensor renormalization group}}, \href{https://doi.org/10.1103/PhysRevD.90.074503}{\emph{Phys. Rev. D} {\bfseries 90} (2014) 074503} [\href{https://arxiv.org/abs/1408.0897}{{\ttfamily 1408.0897}}].

\bibitem{Banuls:2015sta}
M.~C. Ba\~nuls, K.~Cichy, J.~I. Cirac, K.~Jansen and H.~Saito, \emph{{Thermal evolution of the Schwinger model with Matrix Product Operators}}, \href{https://doi.org/10.1103/PhysRevD.92.034519}{\emph{Phys. Rev. D} {\bfseries 92} (2015) 034519} [\href{https://arxiv.org/abs/1505.00279}{{\ttfamily 1505.00279}}].

\bibitem{Buyens:2015tea}
B.~Buyens, J.~Haegeman, H.~Verschelde, F.~Verstraete and K.~Van~Acoleyen, \emph{{Confinement and string breaking for QED$_2$ in the Hamiltonian picture}}, \href{https://doi.org/10.1103/PhysRevX.6.041040}{\emph{Phys. Rev. X} {\bfseries 6} (2016) 041040} [\href{https://arxiv.org/abs/1509.00246}{{\ttfamily 1509.00246}}].

\bibitem{Banuls:2016lkq}
M.~C. Ba\~nuls, K.~Cichy, K.~Jansen and H.~Saito, \emph{{Chiral condensate in the Schwinger model with Matrix Product Operators}}, \href{https://doi.org/10.1103/PhysRevD.93.094512}{\emph{Phys. Rev. D} {\bfseries 93} (2016) 094512} [\href{https://arxiv.org/abs/1603.05002}{{\ttfamily 1603.05002}}].

\bibitem{Buyens:2017crb}
B.~Buyens, S.~Montangero, J.~Haegeman, F.~Verstraete and K.~Van~Acoleyen, \emph{{Finite-representation approximation of lattice gauge theories at the continuum limit with tensor networks}}, \href{https://doi.org/10.1103/PhysRevD.95.094509}{\emph{Phys. Rev. D} {\bfseries 95} (2017) 094509} [\href{https://arxiv.org/abs/1702.08838}{{\ttfamily 1702.08838}}].

\bibitem{Ercolessi:2017jbi}
E.~Ercolessi, P.~Facchi, G.~Magnifico, S.~Pascazio and F.~V. Pepe, \emph{{Phase Transitions in $Z_{n}$ Gauge Models: Towards Quantum Simulations of the Schwinger-Weyl QED}}, \href{https://doi.org/10.1103/PhysRevD.98.074503}{\emph{Phys. Rev. D} {\bfseries 98} (2018) 074503} [\href{https://arxiv.org/abs/1705.11047}{{\ttfamily 1705.11047}}].

\bibitem{Funcke:2019zna}
L.~Funcke, K.~Jansen and S.~K\"uhn, \emph{{Topological vacuum structure of the Schwinger model with matrix product states}}, \href{https://doi.org/10.1103/PhysRevD.101.054507}{\emph{Phys. Rev. D} {\bfseries 101} (2020) 054507} [\href{https://arxiv.org/abs/1908.00551}{{\ttfamily 1908.00551}}].

\bibitem{Magnifico:2019kyj}
G.~Magnifico, M.~Dalmonte, P.~Facchi, S.~Pascazio, F.~V. Pepe and E.~Ercolessi, \emph{{Real Time Dynamics and Confinement in the $\mathbb{Z}_{n}$ Schwinger-Weyl lattice model for 1+1 QED}}, \href{https://doi.org/10.22331/q-2020-06-15-281}{\emph{Quantum} {\bfseries 4} (2020) 281} [\href{https://arxiv.org/abs/1909.04821}{{\ttfamily 1909.04821}}].

\bibitem{Okuda:2022hsq}
T.~Okuda, \emph{{Schwinger model on an interval: Analytic results and DMRG}}, \href{https://doi.org/10.1103/PhysRevD.107.054506}{\emph{Phys. Rev. D} {\bfseries 107} (2023) 054506} [\href{https://arxiv.org/abs/2210.00297}{{\ttfamily 2210.00297}}].

\bibitem{Honda:2022edn}
M.~Honda, E.~Itou and Y.~Tanizaki, \emph{{DMRG study of the higher-charge Schwinger model and its \textquoteright{}t Hooft anomaly}}, \href{https://doi.org/10.1007/JHEP11(2022)141}{\emph{JHEP} {\bfseries 11} (2022) 141} [\href{https://arxiv.org/abs/2210.04237}{{\ttfamily 2210.04237}}].

\bibitem{Angelides:2023bme}
T.~Angelides, L.~Funcke, K.~Jansen and S.~K\"uhn, \emph{{Computing the mass shift of Wilson and staggered fermions in the lattice Schwinger model with matrix product states}}, \href{https://doi.org/10.1103/PhysRevD.108.014516}{\emph{Phys. Rev. D} {\bfseries 108} (2023) 014516} [\href{https://arxiv.org/abs/2303.11016}{{\ttfamily 2303.11016}}].

\bibitem{Martinez:2016yna}
E.~A. Martinez et~al., \emph{{Real-time dynamics of lattice gauge theories with a few-qubit quantum computer}}, \href{https://doi.org/10.1038/nature18318}{\emph{Nature} {\bfseries 534} (2016) 516} [\href{https://arxiv.org/abs/1605.04570}{{\ttfamily 1605.04570}}].

\bibitem{Muschik:2016tws}
C.~Muschik, M.~Heyl, E.~Martinez, T.~Monz, P.~Schindler, B.~Vogell et~al., \emph{{U(1) Wilson lattice gauge theories in digital quantum simulators}}, \href{https://doi.org/10.1088/1367-2630/aa89ab}{\emph{New J. Phys.} {\bfseries 19} (2017) 103020} [\href{https://arxiv.org/abs/1612.08653}{{\ttfamily 1612.08653}}].

\bibitem{Klco:2018kyo}
N.~Klco, E.~F. Dumitrescu, A.~J. McCaskey, T.~D. Morris, R.~C. Pooser, M.~Sanz et~al., \emph{{Quantum-classical computation of Schwinger model dynamics using quantum computers}}, \href{https://doi.org/10.1103/PhysRevA.98.032331}{\emph{Phys. Rev. A} {\bfseries 98} (2018) 032331} [\href{https://arxiv.org/abs/1803.03326}{{\ttfamily 1803.03326}}].

\bibitem{Kokail:2018eiw}
C.~Kokail et~al., \emph{{Self-verifying variational quantum simulation of lattice models}}, \href{https://doi.org/10.1038/s41586-019-1177-4}{\emph{Nature} {\bfseries 569} (2019) 355} [\href{https://arxiv.org/abs/1810.03421}{{\ttfamily 1810.03421}}].

\bibitem{Surace:2019dtp}
F.~M. Surace, P.~P. Mazza, G.~Giudici, A.~Lerose, A.~Gambassi and M.~Dalmonte, \emph{{Lattice gauge theories and string dynamics in Rydberg atom quantum simulators}}, \href{https://doi.org/10.1103/PhysRevX.10.021041}{\emph{Phys. Rev. X} {\bfseries 10} (2020) 021041} [\href{https://arxiv.org/abs/1902.09551}{{\ttfamily 1902.09551}}].

\bibitem{Chakraborty:2020uhf}
B.~Chakraborty, M.~Honda, T.~Izubuchi, Y.~Kikuchi and A.~Tomiya, \emph{{Classically emulated digital quantum simulation of the Schwinger model with a topological term via adiabatic state preparation}}, \href{https://doi.org/10.1103/PhysRevD.105.094503}{\emph{Phys. Rev. D} {\bfseries 105} (2022) 094503} [\href{https://arxiv.org/abs/2001.00485}{{\ttfamily 2001.00485}}].

\bibitem{Kharzeev:2020kgc}
D.~E. Kharzeev and Y.~Kikuchi, \emph{{Real-time chiral dynamics from a digital quantum simulation}}, \href{https://doi.org/10.1103/PhysRevResearch.2.023342}{\emph{Phys. Rev. Res.} {\bfseries 2} (2020) 023342} [\href{https://arxiv.org/abs/2001.00698}{{\ttfamily 2001.00698}}].

\bibitem{Rajput:2021khs}
A.~Rajput, A.~Roggero and N.~Wiebe, \emph{{Hybridized Methods for Quantum Simulation in the Interaction Picture}}, \href{https://doi.org/10.22331/q-2022-08-17-780}{\emph{Quantum} {\bfseries 6} (2022) 780} [\href{https://arxiv.org/abs/2109.03308}{{\ttfamily 2109.03308}}].

\bibitem{Thompson:2021eze}
S.~Thompson and G.~Siopsis, \emph{{Quantum computation of phase transition in the massive Schwinger model}}, \href{https://doi.org/10.1088/2058-9565/ac5f5a}{\emph{Quantum Sci. Technol.} {\bfseries 7} (2022) 035001} [\href{https://arxiv.org/abs/2110.13046}{{\ttfamily 2110.13046}}].

\bibitem{Yamamoto:2021vxp}
A.~Yamamoto, \emph{{Quantum variational approach to lattice gauge theory at nonzero density}}, \href{https://doi.org/10.1103/PhysRevD.104.014506}{\emph{Phys. Rev. D} {\bfseries 104} (2021) 014506} [\href{https://arxiv.org/abs/2104.10669}{{\ttfamily 2104.10669}}].

\bibitem{Honda:2021ovk}
M.~Honda, E.~Itou, Y.~Kikuchi and Y.~Tanizaki, \emph{{Negative string tension of a higher-charge Schwinger model via digital quantum simulation}}, \href{https://doi.org/10.1093/ptep/ptac007}{\emph{PTEP} {\bfseries 2022} (2022) 033B01} [\href{https://arxiv.org/abs/2110.14105}{{\ttfamily 2110.14105}}].

\bibitem{deJong:2021wsd}
W.~A. de~Jong, K.~Lee, J.~Mulligan, M.~P\l{}osko\'n, F.~Ringer and X.~Yao, \emph{{Quantum simulation of nonequilibrium dynamics and thermalization in the Schwinger model}}, \href{https://doi.org/10.1103/PhysRevD.106.054508}{\emph{Phys. Rev. D} {\bfseries 106} (2022) 054508} [\href{https://arxiv.org/abs/2106.08394}{{\ttfamily 2106.08394}}].

\bibitem{Honda:2021aum}
M.~Honda, E.~Itou, Y.~Kikuchi, L.~Nagano and T.~Okuda, \emph{{Classically emulated digital quantum simulation for screening and confinement in the Schwinger model with a topological term}}, \href{https://doi.org/10.1103/PhysRevD.105.014504}{\emph{Phys. Rev. D} {\bfseries 105} (2022) 014504} [\href{https://arxiv.org/abs/2105.03276}{{\ttfamily 2105.03276}}].

\bibitem{Rajput:2021trn}
A.~Rajput, A.~Roggero and N.~Wiebe, \emph{{Quantum error correction with gauge symmetries}}, \href{https://doi.org/10.1038/s41534-023-00706-8}{\emph{npj Quantum Inf.} {\bfseries 9} (2023) 41} [\href{https://arxiv.org/abs/2112.05186}{{\ttfamily 2112.05186}}].

\bibitem{Nguyen:2021hyk}
N.~H. Nguyen, M.~C. Tran, Y.~Zhu, A.~M. Green, C.~H. Alderete, Z.~Davoudi et~al., \emph{{Digital Quantum Simulation of the Schwinger Model and Symmetry Protection with Trapped Ions}}, \href{https://doi.org/10.1103/PRXQuantum.3.020324}{\emph{PRX Quantum} {\bfseries 3} (2022) 020324} [\href{https://arxiv.org/abs/2112.14262}{{\ttfamily 2112.14262}}].

\bibitem{Cheng:2022jnw}
Y.~Cheng, S.~Liu, W.~Zheng, P.~Zhang and H.~Zhai, \emph{{Tunable Confinement-Deconfinement Transition in an Ultracold-Atom Quantum Simulator}}, \href{https://doi.org/10.1103/PRXQuantum.3.040317}{\emph{PRX Quantum} {\bfseries 3} (2022) 040317} [\href{https://arxiv.org/abs/2204.06586}{{\ttfamily 2204.06586}}].

\bibitem{Tomiya:2022chr}
A.~Tomiya, \emph{{Schwinger model at finite temperature and density with beta VQE}},  \href{https://arxiv.org/abs/2205.08860}{{\ttfamily 2205.08860}}.

\bibitem{Nagano:2023uaq}
L.~Nagano, A.~Bapat and C.~W. Bauer, \emph{{Quench dynamics of the Schwinger model via variational quantum algorithms}}, \href{https://doi.org/10.1103/PhysRevD.108.034501}{\emph{Phys. Rev. D} {\bfseries 108} (2023) 034501} [\href{https://arxiv.org/abs/2302.10933}{{\ttfamily 2302.10933}}].

\bibitem{Ikeda:2023zil}
K.~Ikeda, D.~E. Kharzeev, R.~Meyer and S.~Shi, \emph{{Detecting the critical point through entanglement in the Schwinger model}}, \href{https://doi.org/10.1103/PhysRevD.108.L091501}{\emph{Phys. Rev. D} {\bfseries 108} (2023) L091501} [\href{https://arxiv.org/abs/2305.00996}{{\ttfamily 2305.00996}}].

\bibitem{Sakamoto:2023cxs}
K.~Sakamoto, H.~Morisaki, J.~Haruna, E.~Itou, K.~Fujii and K.~Mitarai, \emph{{End-to-end complexity for simulating the Schwinger model on quantum computers}}, \href{https://doi.org/10.22331/q-2024-09-17-1474}{\emph{Quantum} {\bfseries 8} (2024) 1474} [\href{https://arxiv.org/abs/2311.17388}{{\ttfamily 2311.17388}}].

\bibitem{Farrell:2023fgd}
R.~C. Farrell, M.~Illa, A.~N. Ciavarella and M.~J. Savage, \emph{{Scalable Circuits for Preparing Ground States on Digital Quantum Computers: The Schwinger Model Vacuum on 100 Qubits}}, \href{https://doi.org/10.1103/PRXQuantum.5.020315}{\emph{PRX Quantum} {\bfseries 5} (2024) 020315} [\href{https://arxiv.org/abs/2308.04481}{{\ttfamily 2308.04481}}].

\bibitem{Farrell:2024fit}
R.~C. Farrell, M.~Illa, A.~N. Ciavarella and M.~J. Savage, \emph{{Quantum simulations of hadron dynamics in the Schwinger model using 112 qubits}}, \href{https://doi.org/10.1103/PhysRevD.109.114510}{\emph{Phys. Rev. D} {\bfseries 109} (2024) 114510} [\href{https://arxiv.org/abs/2401.08044}{{\ttfamily 2401.08044}}].

\bibitem{Ghim:2024pxe}
D.~Ghim and M.~Honda, \emph{{Digital Quantum Simulation for Spectroscopy of Schwinger Model}}, \href{https://doi.org/10.22323/1.453.0213}{\emph{PoS} {\bfseries LATTICE2023} (2024) 213} [\href{https://arxiv.org/abs/2404.14788}{{\ttfamily 2404.14788}}].

\bibitem{Kaikov:2024acw}
O.~Kaikov, T.~Saporiti, V.~Sazonov and M.~Tamaazousti, \emph{{Phase Diagram of the Schwinger Model by Adiabatic Preparation of States on a Quantum Simulator}},  \href{https://arxiv.org/abs/2407.09224}{{\ttfamily 2407.09224}}.

\bibitem{Guo:2024tnb}
Y.~Guo, T.~Angelides, K.~Jansen and S.~K\"uhn, \emph{{Concurrent VQE for Simulating Excited States of the Schwinger Model}},  \href{https://arxiv.org/abs/2407.15629}{{\ttfamily 2407.15629}}.

\bibitem{Araz:2024bgg}
J.~Y. Araz, S.~Bhowmick, M.~Grau, T.~J. McEntire and F.~Ringer, \emph{{State preparation of lattice field theories using quantum optimal control}},  \href{https://arxiv.org/abs/2407.17556}{{\ttfamily 2407.17556}}.

\bibitem{Li:2024jlo}
X.-W. Li, F.~Li, J.~Zhuang and M.-H. Yung, \emph{{Simulating the Schwinger Model with a Regularized Variational Quantum Imaginary Time Evolution}},  \href{https://arxiv.org/abs/2409.13510}{{\ttfamily 2409.13510}}.

\bibitem{2011AnPhy.326...96S}
U.~{Schollw{\"o}ck}, \emph{{The density-matrix renormalization group in the age of matrix product states}}, \href{https://doi.org/10.1016/j.aop.2010.09.012}{\emph{Annals of Physics} {\bfseries 326} (2011) 96} [\href{https://arxiv.org/abs/1008.3477}{{\ttfamily 1008.3477}}].

\bibitem{Zauner-Stauber:2017eqw}
V.~Zauner-Stauber, L.~Vanderstraeten, M.~T. Fishman, F.~Verstraete and J.~Haegeman, \emph{{Variational optimization algorithms for uniform matrix product states}}, \href{https://doi.org/10.1103/PhysRevB.97.045145}{\emph{Phys. Rev. B} {\bfseries 97} (2018) 045145} [\href{https://arxiv.org/abs/1701.07035}{{\ttfamily 1701.07035}}].

\bibitem{Kogut:1974ag}
J.~B. Kogut and L.~Susskind, \emph{{Hamiltonian Formulation of Wilson's Lattice Gauge Theories}}, \href{https://doi.org/10.1103/PhysRevD.11.395}{\emph{Phys. Rev. D} {\bfseries 11} (1975) 395}.

\bibitem{White:1992zz}
S.~R. White, \emph{{Density matrix formulation for quantum renormalization groups}}, \href{https://doi.org/10.1103/PhysRevLett.69.2863}{\emph{Phys. Rev. Lett.} {\bfseries 69} (1992) 2863}.

\bibitem{McCulloch:2008aun}
I.~P. McCulloch, \emph{{Infinite size density matrix renormalization group, revisited}},  \href{https://arxiv.org/abs/0804.2509}{{\ttfamily 0804.2509}}.

\bibitem{Vidal:2006ofj}
G.~Vidal, \emph{{Classical simulation of infinite-size quantum lattice systems in one spatial dimension}}, \href{https://doi.org/10.1103/PhysRevLett.98.070201}{\emph{Phys. Rev. Lett.} {\bfseries 98} (2007) 070201} [\href{https://arxiv.org/abs/cond-mat/0605597}{{\ttfamily cond-mat/0605597}}].

\bibitem{Haegeman:2011zz}
J.~Haegeman, J.~I. Cirac, T.~J. Osborne, I.~Pizorn, H.~Verschelde and F.~Verstraete, \emph{{Time-Dependent Variational Principle for Quantum Lattices}}, \href{https://doi.org/10.1103/PhysRevLett.107.070601}{\emph{Phys. Rev. Lett.} {\bfseries 107} (2011) 070601} [\href{https://arxiv.org/abs/1103.0936}{{\ttfamily 1103.0936}}].

\bibitem{2014arXiv1408.5056H}
J.~{Haegeman}, C.~{Lubich}, I.~{Oseledets}, B.~{Vandereycken} and F.~{Verstraete}, \emph{{Unifying time evolution and optimization with matrix product states}}, \href{https://doi.org/10.48550/arXiv.1408.5056}{\emph{arXiv e-prints} (2014) arXiv:1408.5056} [\href{https://arxiv.org/abs/1408.5056}{{\ttfamily 1408.5056}}].

\bibitem{Coleman:1976uz}
S.~R. Coleman, \emph{{More About the Massive Schwinger Model}}, \href{https://doi.org/10.1016/0003-4916(76)90280-3}{\emph{Annals Phys.} {\bfseries 101} (1976) 239}.

\bibitem{Hamer:1982mx}
C.~J. Hamer, J.~B. Kogut, D.~P. Crewther and M.~M. Mazzolini, \emph{{The Massive Schwinger Model on a Lattice: Background Field, Chiral Symmetry and the String Tension}}, \href{https://doi.org/10.1016/0550-3213(82)90229-2}{\emph{Nucl. Phys. B} {\bfseries 208} (1982) 413}.

\bibitem{Vanhecke:2019pez}
B.~Vanhecke, J.~Haegeman, K.~Van~Acoleyen, L.~Vanderstraeten and F.~Verstraete, \emph{{Scaling Hypothesis for Matrix Product States}}, \href{https://doi.org/10.1103/PhysRevLett.123.250604}{\emph{Phys. Rev. Lett.} {\bfseries 123} (2019) 250604} [\href{https://arxiv.org/abs/1907.08603}{{\ttfamily 1907.08603}}].

\bibitem{Vanhecke:2021noi}
B.~Vanhecke, F.~Verstraete and K.~Van~Acoleyen, \emph{{Entanglement scaling for \ensuremath{\lambda}\ensuremath{\phi}24}}, \href{https://doi.org/10.1103/PhysRevD.106.L071501}{\emph{Phys. Rev. D} {\bfseries 106} (2022) L071501} [\href{https://arxiv.org/abs/2104.10564}{{\ttfamily 2104.10564}}].

\bibitem{ArguelloCruz:2024xzi}
E.~Arguello~Cruz, G.~Tarnopolsky and Y.~Xin, \emph{{Precision study of the massive Schwinger model near quantum criticality}},  \href{https://arxiv.org/abs/2412.01902}{{\ttfamily 2412.01902}}.

\bibitem{Banks:1975gq}
T.~Banks, L.~Susskind and J.~B. Kogut, \emph{{Strong Coupling Calculations of Lattice Gauge Theories: (1+1)-Dimensional Exercises}}, \href{https://doi.org/10.1103/PhysRevD.13.1043}{\emph{Phys. Rev. D} {\bfseries 13} (1976) 1043}.

\bibitem{Jordan:1928wi}
P.~Jordan and E.~P. Wigner, \emph{{About the Pauli exclusion principle}}, \href{https://doi.org/10.1007/BF01331938}{\emph{Z. Phys.} {\bfseries 47} (1928) 631}.

\bibitem{Dempsey:2022nys}
R.~Dempsey, I.~R. Klebanov, S.~S. Pufu and B.~Zan, \emph{{Discrete chiral symmetry and mass shift in the lattice Hamiltonian approach to the Schwinger model}}, \href{https://doi.org/10.1103/PhysRevResearch.4.043133}{\emph{Phys. Rev. Res.} {\bfseries 4} (2022) 043133} [\href{https://arxiv.org/abs/2206.05308}{{\ttfamily 2206.05308}}].

\bibitem{Berruto:1999ga}
F.~Berruto, G.~Grignani, G.~W. Semenoff and P.~Sodano, \emph{{On the correspondence between the strongly coupled two flavor lattice Schwinger model and the Heisenberg antiferromagnetic chain}}, \href{https://doi.org/10.1006/aphy.1999.5934}{\emph{Annals Phys.} {\bfseries 275} (1999) 254} [\href{https://arxiv.org/abs/hep-th/9901142}{{\ttfamily hep-th/9901142}}].

\bibitem{VanAcoleyen:2014suo}
K.~Van~Acoleyen, B.~Buyens, J.~Haegeman and F.~Verstraete, \emph{{Matrix product states for Hamiltonian lattice gauge theories}}, \href{https://doi.org/10.22323/1.214.0308}{\emph{PoS} {\bfseries LATTICE2014} (2014) 308} [\href{https://arxiv.org/abs/1411.0020}{{\ttfamily 1411.0020}}].

\bibitem{Buyens:2016ecr}
B.~Buyens, F.~Verstraete and K.~Van~Acoleyen, \emph{{Hamiltonian simulation of the Schwinger model at finite temperature}}, \href{https://doi.org/10.1103/PhysRevD.94.085018}{\emph{Phys. Rev. D} {\bfseries 94} (2016) 085018} [\href{https://arxiv.org/abs/1606.03385}{{\ttfamily 1606.03385}}].

\bibitem{Buyens:2016hhu}
B.~Buyens, J.~Haegeman, F.~Hebenstreit, F.~Verstraete and K.~Van~Acoleyen, \emph{{Real-time simulation of the Schwinger effect with Matrix Product States}}, \href{https://doi.org/10.1103/PhysRevD.96.114501}{\emph{Phys. Rev. D} {\bfseries 96} (2017) 114501} [\href{https://arxiv.org/abs/1612.00739}{{\ttfamily 1612.00739}}].

\bibitem{Zauner:2014iea}
V.~Zauner, D.~Draxler, L.~Vanderstraeten, M.~Degroote, J.~Haegeman, M.~M. Rams et~al., \emph{{Transfer Matrices and Excitations with Matrix Product States}}, \href{https://doi.org/10.1088/1367-2630/17/5/053002}{\emph{New J. Phys.} {\bfseries 17} (2015) 053002} [\href{https://arxiv.org/abs/1408.5140}{{\ttfamily 1408.5140}}].

\bibitem{ornstein1914accidental}
L.~S. Ornstein and F.~Zernike, \emph{Accidental deviations of density and opalescence at the critical point of a single substance}, {\emph{Proc. Akad. Sci.} {\bfseries 17} (1914) 793}.

\bibitem{Kennedy1991OrnsteinZernikeDI}
T.~Kennedy, \emph{Ornstein-zernike decay in the ground state of the quantum ising model in a strong transverse field}, {\emph{Communications in Mathematical Physics} {\bfseries 137} (1991) 599}.

\bibitem{Cardy:1996xt}
J.~L. Cardy, \emph{{Scaling and renormalization in statistical physics}}. {Cambridge University Press, Cambridge, England}, 1996.

\bibitem{2018PhRvX...8d1033R}
M.~M. {Rams}, P.~{Czarnik} and L.~{Cincio}, \emph{{Precise Extrapolation of the Correlation Function Asymptotics in Uniform Tensor Network States with Application to the Bose-Hubbard and XXZ Models}}, \href{https://doi.org/10.1103/PhysRevX.8.041033}{\emph{Physical Review X} {\bfseries 8} (2018) 041033} [\href{https://arxiv.org/abs/1801.08554}{{\ttfamily 1801.08554}}].

\bibitem{10.5555/1403886}
W.~H. Press, S.~A. Teukolsky, W.~T. Vetterling and B.~P. Flannery, \emph{Numerical Recipes 3rd Edition: The Art of Scientific Computing}. Cambridge University Press, USA, 3~ed., 2007.

\bibitem{Calabrese:2004eu}
P.~Calabrese and J.~L. Cardy, \emph{{Entanglement entropy and quantum field theory}}, \href{https://doi.org/10.1088/1742-5468/2004/06/P06002}{\emph{J. Stat. Mech.} {\bfseries 0406} (2004) P06002} [\href{https://arxiv.org/abs/hep-th/0405152}{{\ttfamily hep-th/0405152}}].

\bibitem{Gepner:1984au}
D.~Gepner, \emph{{Nonabelian Bosonization and Multiflavor {QED} and {QCD} in Two-dimensions}}, \href{https://doi.org/10.1016/0550-3213(85)90458-4}{\emph{Nucl. Phys. B} {\bfseries 252} (1985) 481}.

\bibitem{Affleck:1985wa}
I.~Affleck, \emph{{On the Realization of Chiral Symmetry in (1+1)-dimensions}}, \href{https://doi.org/10.1016/0550-3213(86)90168-9}{\emph{Nucl. Phys. B} {\bfseries 265} (1986) 448}.

\bibitem{Dempsey:2023gib}
R.~Dempsey, I.~R. Klebanov, S.~S. Pufu, B.~T. S\o{}gaard and B.~Zan, \emph{{Phase Diagram of the Two-Flavor Schwinger Model at Zero Temperature}}, \href{https://doi.org/10.1103/PhysRevLett.132.031603}{\emph{Phys. Rev. Lett.} {\bfseries 132} (2024) 031603} [\href{https://arxiv.org/abs/2305.04437}{{\ttfamily 2305.04437}}].

\bibitem{Dempsey:2023fvm}
R.~Dempsey, I.~R. Klebanov, S.~S. Pufu and B.~T. S\o{}gaard, \emph{{Lattice Hamiltonian for adjoint QCD$_{2}$}}, \href{https://doi.org/10.1007/JHEP08(2024)009}{\emph{JHEP} {\bfseries 08} (2024) 009} [\href{https://arxiv.org/abs/2311.09334}{{\ttfamily 2311.09334}}].

\end{thebibliography}\endgroup

\end{document}